# What Are People Tweeting about Zika? An Exploratory Study Concerning Symptoms, Treatment, Transmission, and Prevention

Corresponding Author
Michele Miller MS
Wright State University
Biological Sciences
3640 Colonel Glenn Hwy, Dayton, OH 45431
Phone (574)261-3969
Fax (937)775-3320
miller.1232@wright.edu

Dr. Tanvi Banerjee PhD
Wright State University
Computer Science and Engineering
3640 Colonel Glenn Hwy, Dayton, OH 45431

Roopteja Muppalla
Wright State University
Computer Science and Engineering
3640 Colonel Glenn Hwy, Dayton, OH 45431

Dr. William Romine PhD
Wright State University
Biological Sciences
3640 Colonel Glenn Hwy, Dayton, OH 45431

Dr. Amit Sheth PhD
Wright State University
Computer Science and Engineering
3640 Colonel Glenn Hwy, Dayton, OH 45431

**Abstract.**
*Background* So far an exploratory study concerning what people are tweeting about Zika has not been performed.
*Objective* The purpose of this study was to do a dataset distribution analysis, a classification performance analysis, and a topical analysis concerning what people are tweeting about four disease characteristics: symptoms, transmission, prevention, and treatment.
*Methods* A combination of natural language processing and machine learning techniques were used to determine what people are tweeting about Zika. Specifically, a two-stage classifier system was built to find relevant tweets on Zika, and then categorize these into the four disease categories. Tweets in each disease category were then examined using latent dirichlet allocation (LDA) to determine the five main tweet topics for each disease characteristic.
*Results* 1,234,605 tweets were collected. Tweets by males and females were similar (28% and 23% respectively). The classifier performed well on the training and test data for relevancy (F=0.87 and 0.99 respectively) and disease characteristics (F=0.79 and 0.90 respectively). Five topics for each category were found and discussed with a focus on the symptoms category.
*Conclusion* Through this process, we demonstrate how misinformation can be discovered so that public health officials can respond to the tweets with misinformation.

# Introduction

The 2014/2015 Ebola outbreak caused fear and misinformation to spread wildly across the globe. It was shown that the spread of misinformation led to deaths due to improper practice of appropriate preventative measures [1].

Experts at the Center for Disease Control (CDC) and the World Health Organization (WHO) admit that they botched the response for Ebola by not responding to the threat sooner [2]. One year after the Ebola outbreak ended, the Zika outbreak started and also caused fear and misinformation to spread. In the recent years, citizen sensing has picked up greatly with the rise of mobile device popularity, as well as with the rise in social media sites such as Facebook and Twitter. The idea with citizen sensing is that citizens play the role of sensors in the environment [3]; providing information regarding healthcare issues such as disease outbreaks in case of Ebola and Zika [4].

Big social data eliminate the time lag caused by traditional survey based methods, allowing for studying public opinions on issues while addressing privacy concerns of users by studying collective public behavior on specific issues. In particular, public opinion mining has been studied in the past for exploration of public views on important social issues such as gender-based violence [5], as well as to mine health related beliefs [6-7].

With respect to Zika, Twitter has served as a source of misinformation. To counter, the CDC has been responding with correct information. For example, one user tweeted "*Apparently Florida is immune to the Zika virus*" while the CDC has tweeted about Zika in Florida several times including this tweet "*Updated: CDC travel and testing recommendations for Miami-Dade county b/c of continued local #Zika transmission*". Another common misconception is found in this re-tweet "*RT @user: I saw the Zika virus has made its way to Houston. It's really only bad for you if you are pregnant .. *". However, while Zika is typically mild to unnoticeable in adults, it can cause health issues in some adults as the CDC explains in this tweet "*Symptoms of Guillain-Barre syndrome include weakness in arms and legs. GBS (Guillain-Barre syndrome) is linked w/ #Zika. Learn more. bit.ly/2dqL7Dv*". One more common misconception is that "*It won't do any good 2 attempt 2 prevent Zika Virus; instead, devote all funds 2 researching a vaccine, or anecdote*". Firstly, assuming that the tweet meant "antidote" instead of "anecdote," the statement is still incorrect since antidotes are a substance that stops the harmful effects of poison. This tweet puts forth a fatalistic attitude towards Zika prevention in its indication that there is nothing citizens can do to prevent Zika. The CDC responds to this by posting several tweets about what the public can do to prevent Zika infection such as "*Treating your clothing & gear w/ permethrin can help prevent mosquito bites. Learn more ways to #ZapZika*" and

"*Prevent #Zika spread after travel. Use condoms: 6 months after travel for men, 8 weeks for women*".

In this exploratory study, a combination of natural language processing and machine learning techniques were used to determine what information about Zika symptoms, transmission, prevention, and treatment people were discussing using tweets. Specifically, a two-stage classifier system was built that was used to find relevant tweets on Zika, and then categorize those into four disease categories: symptoms, transmission, prevention, and treatment (Figure 1). This information could then be used by health professionals such as people at the CDC and WHO to know what information the public does and does not know about Zika. Such a system may help inform them on what they need to include in messages to the public, as well as target specific user groups to prevent the spread of misinformation.

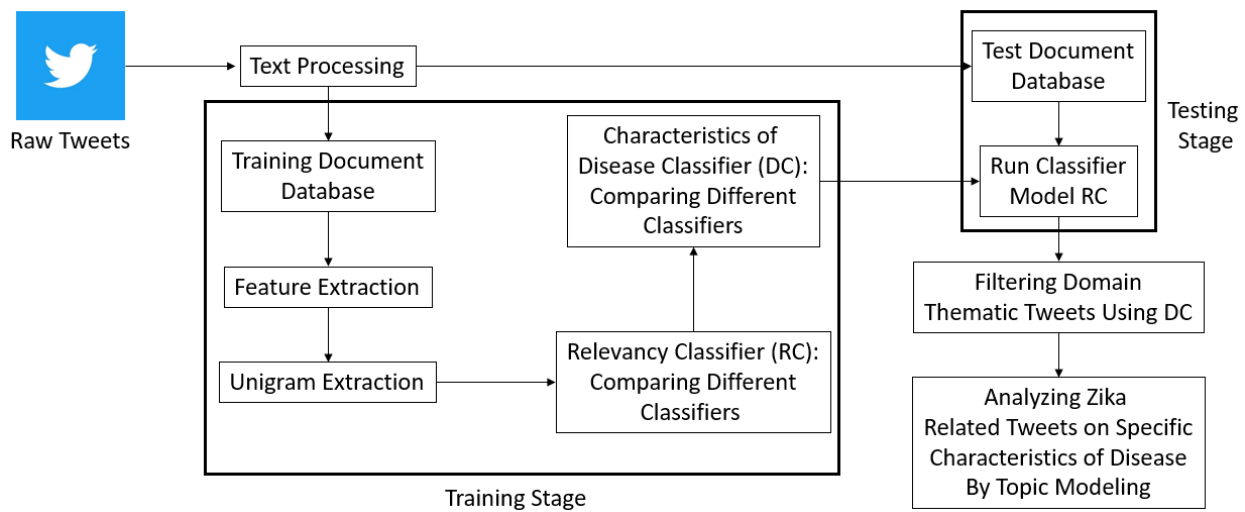

Figure 1. Block diagram of the pragmatic function-oriented content retrieval using a hierarchical supervised classification technique, followed by deeper analysis for characteristics of disease content.

**Zika**

Zika belongs to the Flaviviridae family of viruses [8]. This family contains a few arboviruses that are very important clinically, such as dengue virus, yellow fever virus, West Nile Virus, tick-borne encephalitis virus, and Japanese encephalitis virus [8]. Prior to 2007 there had only been 14 confirmed cases of Zika; however, many people do not even realize they are sick from Zika, let alone need to go to the hospital and death due to Zika is extremely rare [9]. The Zika virus usually only causes mild symptoms such as a headache, rash, fever, conjunctivitis, and joint pain which can last from a few days to a week after being infected. These symptoms are similar to Dengue and Chikungunya, which are also spread by the same mosquito as Zika. The incubation period is unknown at this time but suspected to be a few days to a week. The virus typically remains in the blood for about a week, but has been found later in some people. People at risk of getting Zika include people that travel or live in an area where Zika has been found and has been bit by a mosquito, and partners of men that have traveled to an area with the Zika virus. In the next several months the countries with active cases is expected to increase [9].

Complications such as Guillain-Barré syndrome, microcephaly, and maculopathy may occur in infants [8]. In Brazil, the number of infants with microcephaly increased 20 times after the start of the Zika virus epidemic. Fetuses and infants have also been found to have cerebral calcification and retinal abnormalities such as macular pigment mottling, macular neuroretinal atrophy, optic nerve hypoplasia, chorioretinal atrophy, and foveal reflex loss. Since this is the first outbreak of Zika associated with these defects, management is still an important challenge.

There are three main ways to get Zika: (i) being bitten by an infected *Aedes* mosquito, (ii) through sexual contact, and (iii) from mother to fetus [9]. The *Aedes* mosquito aggressively bites during the daytime, but can also bite at night. Mosquitos get the virus by feeding on someone with the virus and then spread it by feeding on other people. An infected mother can pass the virus to a newborn during pregnancy. It has been shown that the virus can somehow cross the placenta. As of yet, there is no evidence of a mother passing the virus to her newborn through breastfeeding. There have been cases of an infected man spreading the virus to his partner during intercourse. It has been found that the virus persists in semen longer than blood and that the virus can be spread before, during, and after, the man has symptoms. There have been no confirmed cases of someone in the US getting Zika after receiving a blood transfusion as of February, 1, 2016. However, there have been reports of people in Brazil getting Zika from a blood transfusion [9].

A person is most likely protected from future infections once already infected [9]. There is currently no medicine or vaccine to treat the Zika virus. Experts suggest rest, plenty of fluids, and acetaminophen or paracetamol for fever and pain. They say to avoid aspirin and other non-steroidal anti-inflammatory drugs until it is confirmed to be Zika and not Dengue. Sexual transmission can be prevented by abstaining or using condoms. Infection by mosquito bite can be prevented by wearing long-sleeved shirts and long pants, staying in places with air conditioning, staying in places that have door and window screens, sleeping under a mosquito net, and using insect repellants. People returning from places with Zika should prevent being bitten by a mosquito for three weeks to prevent the spread of the virus to uninfected mosquitoes [9].

**Related Works**

A study about misinformation about Ebola on Twitter found that a lot of misinformation about cures for Ebola was circulating around Twitter [10]. They found that 44% (248) of the tweets about Ebola were retweeted at least once. Of those 248 tweets, 38.3% were scientifically accurate while 58.9% were inaccurate. They also found that most of the tweets containing misinformation were never corrected.

Alvaro et al. [11] used a Twitter streaming application program interface (API) to obtain a random sample of tweets collected over a 12-month period about first-hand experience with cognitive enhancers or selective serotonin reuptake inhibitors. Tweets were categorized by drug and an upper bound of 300 tweets per drug was set to collect a balanced sample with a total of 1548 tweets actually used for the study. One hundred tweets were annotated and used as the ground truth to evaluate other annotators from crowd sourcing. In order to verify agreement between layman annotators and experienced annotators, Cohen's Kappa and Fleiss's Kappa were computed for different questions such as whether the tweet is in English, whether the tweet contains drug mention, etc. These scores are then ranked using Spearman's Rho and Kendall's tau with values of 0.471 and 0.352 respectively. N-grams, hashtags, URLs, were extracted and classifiers were built. Bayesian Generalized Linear Modeling was found to be the best technique for interpretation of the data in this study.

One study sought to understand how people responded to the Ebola outbreak on social media, what type of messages people post, what factors affect these reactions, and patterns within these reactions [4]. At first 2 billion tweets were collected using the keyword 'Ebola' from 90 different languages. After removing duplicates and the languages with less than 100 tweets, the final tweet count is 1,167,539 tweets from 35 languages, out of which 569,888 are geotagged. For the non-geotagged tweets, users' profile locations on Twitter were used to get the location information that was then analyzed. In order to investigate the spatio-temporal properties about the information propagation from the central cities (the cities where the outbreak is reported, in this case, New York and Dallas), they considered the tweets (46,598) which are over 2 days from the starting day of the outbreak in these cities. Results also showed that the first reported incident had more impact and received more attention than any other showing that people pay more attention and react more since it is a new topic. Finally, it showed the significance of social ties in propagating the information by analyzing how the Twitter followers post/retweet based on their followee's tweets.

Another study used Twitter to track 300 million tweets about the spread of influenza from November 2008 to June 2010 [12]. Tweets were annotated as being positive or negative by annotators, with positive being tweets where the person with the flu tweeted or someone around a person with the flu tweeted. All other tweets were considered negative. This positive and negative annotations were considered their influenza corpus. Using a support vector machine (SVM) based classifier, they eliminated tweets that did not actually mention flu patients based on their developed influenza corpus. The tweets were divided into four seasons: winter 2008, summer 2009, winter 2009, and summer 2010 for the test-set. Annotations and estimated values were compared using the Pearson correlation. They found their method performed well during non-news periods (0.89 correlation) but did not perform well during excessive news periods, like during the H1N1 flu epidemic. While these different studies highlighted the utility of using social media to monitor people's thoughts regarding a specific disease outbreak, they did not explore the specific themes within a disease.

In this study we conduct an exploratory analysis of finding the subcategories of discussion topics from Zika related tweets; specifically in four key characteristics of Zika: symptoms, transmission, treatment, and prevention. Using the system described in Figure 1, we address the following research questions:

*R1:* Dataset Distribution Analysis: What proportion of male and female users tweeted about Zika, what were the polarities of the tweets by male and female users, and what were the proportion of tweets that discussed topics related to the different disease characteristics - symptoms, transmission, treatment, and prevention?

*R2:* Classification Performance Analysis: What was the agreement among the annotators who labeled a sample of the data that was used as ground truth in this study, what was the classification performance to detect the tweets relevant to Zika, and how well were the classifiers able to detect between tweets on the different disease characteristics?

*R3:* Topical Analysis: What were the main discussion topics in each of these characteristics that can further inform members at CDC on the biggest concerns or misconceptions regarding Zika virus?

# Methods
**Data Collection**

Tweets are collected between the time 2016-02-24 and 2016-04-27 for a total of 1,234,605 tweets using Twitris 2.0 [13]. The keywords used to collect the tweets were "Zika," "Zika virus," "Zika treatment," and "Zika virus treatment." One thousand four hundred and sixty-seven random tweets were put in the following categories: relevant or non-relevant, and then if they were relevant they were further categorized as symptoms, treatment, transmission, and prevention. These four categories were used because they are characteristics of disease used in any medical journal and by the CDC and WHO and are what is important for scientists and medical professionals to know about a disease.

**Labeling Process/ Data Annotation**

One-thousand four hundred and sixty-seven tweets were annotated as relevant or not, and then by the four subcategories if they were relevant, by three microbiology and immunology experts. Inter-rater reliability was found using Fleiss Kappa [14].

**Preprocessing**

As the initial step, the data were further preprocessed to remove the URL, screen handles (@username), retweet indicators, and non-ascii characters. Data were further normalized by removing capital letters, numbers, punctuations and whitespaces from the tweets. Terms were filtered out to remove single characters like 'd', 'e' which do not convey any meaning about the topics in the corpus and top words like "and", "so", etc were removed For the classification stage, each tweet is represented as a feature vector of the words present in the tweet using unigrams.

**Classification**

Supervised classification techniques including the decision tree (J48), multinomial naive bayes (MNB), Bayesian networks (Bayes Net), SMO (sequential minimal optimization) using SVM, Adaboost, as well as bagging or bootstrapping (Bagging) techniques were implemented on the Zika dataset for a) classifying relevant tweets on zika, and b) if relevant, further categorizing it into the disease characteristics. Supervised techniques rely on labeled data, in this case tweets, that are manually labeled as relevant to Zika virus, as well as the category it belongs to: Zika symptoms, Zika treatment, Zika transmission, and Zika prevention. They "learn" the nature of the tweets in the different groups and subgroups.

The performance of each classifier was assessed using 10-fold cross validation, which is a commonly used method for the evaluation of classification algorithms that diminishes the bias in the estimation of classifier performance [15]. This approach uses the entire dataset for both training and testing, and is especially useful when the manually labeled data set is relatively small. In 10 fold cross-validation, the manually labeled data set is randomly partitioned into 10 equal-sized subsets. The cross-validation process is then repeated 10 times (the folds). Each time, a single subset is retained as the validation data for testing the model, and the remaining 9 subsamples are used as training data. The 10 results from the folds are then averaged to produce a single estimation. The study reports the average of the precision, recall, F-scores, and area under the curve (AUC) calculated by the system. Precision is defined as the number of correctly classified positive examples divided by the number of examples labeled by the system as positive. Recall, also referred to as sensitivity, is defined as the number of correctly classified positive examples divided by the number of positive examples in the manually coded gold standard data. An F-score is a combination (harmonic mean) of precision and recall measures [15]. The area under the curve also evaluates the tradeoff between precision and recall using different threshold values that affect the sensitivity as well as the specificity of the different algorithms. For the ideal classifier the AUC value would be 1, and the AUC for a random predictor would be 0.5, so most classifiers have a value between 0.5 and 1.

**Topic Modeling**

Studies such as Hong [16] have shown the utility of using traditional topic modeling methods like LDA for grouping the themes occurring in short text documents. The basic idea in LDA is that documents (tweets in this case) are represented as random mixtures over hidden topics, where each topic is characterized by a distribution over words that occur the most frequently within that topic [17]. As an example, if person A makes a decision to sit at a particular table T at a conference, the probability of A sitting at T is proportional to the other people sitting at T. Also, as more people join the conference, the probabilities for choosing tables converge. In LDAs the people are the words, and the tables are the topics present in a given dataset. In this study, we use topic modeling for finding the underlying topics in each of the four disease characteristics to find out more about the important issues in each of these categories.

# Results

**Dataset Distribution (addressing R1):**

Overall, 42% of tweets contained a retweet and 85% contained a URL. Tweets by gender were found using the twitter usernames using the genderize API [18]. According to genderize, twenty-eight percent of the tweets were by males, twenty-three percent by females and 41% were by unknown gender.

The polarity of the individual tweets were also found using the sentiment package in R [19] (Table 1).

Table 1. Polarity and proportion of tweets divided in the gender categories.

| Category | All tweets | Male tweets | Female tweets |
| --- | --- | --- | --- |
| Positive | 313,742 (24%) | 90,928 (26%) | 70,171 (25%) |
| Negative | 783,327 (59%) | 220,359 (63%) | 180,276 (63%) |
| Neutral | 137,536 (10%) | 40,166 (11%) | 33,760 (12%) |

The polarity of the tweets between males and females were similar. Although a majority of the tweets are categorized as having negative polarity, the percent of positive tweets was higher than expected. Some examples of tweets that were classified as positive were "*Case report: assoc btw #Zika/teratogenicity strengthened & evidence shows impact on fetus may take time to manifest*", "*RT @NEJM: At recent int'l meeting about #Zika, experts exchanged insights, identified knowledge gaps, and agreed on a plan*", and "*91,387 Cases of Zika Confirmed in Brazil This Year: Brazil has confirmed 91,387 cases of…*". Words such as "strengthened", "agreed", "confirmed" may be why some tweets are being classified as positive.

There was a class imbalance in the categories. Since there is no treatment for Zika, not many people tweeted about it. Transmission and prevention tweets were high since they were the most talked about topics concerning Zika. Sample tweets from these different categories as provided in Table 3.

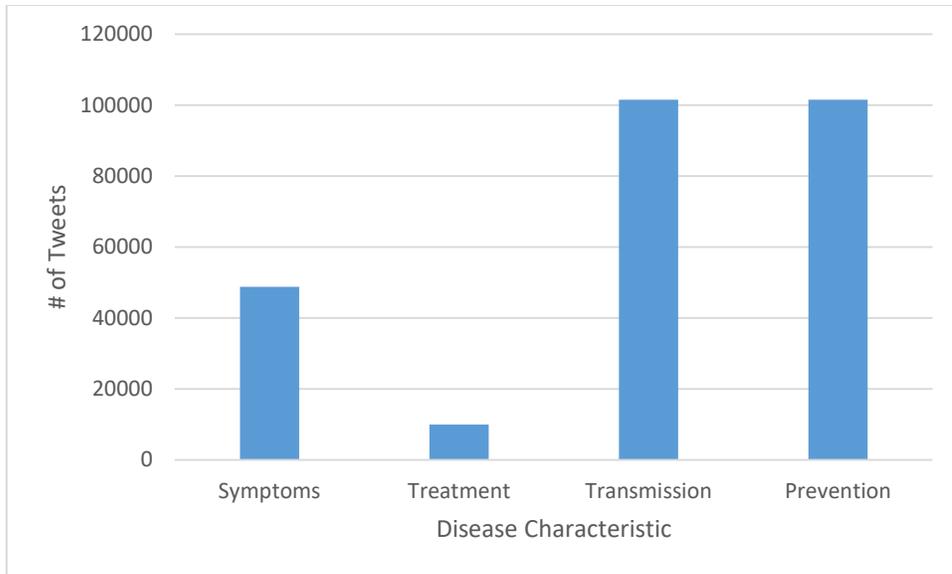

Figure 2. Number of tweets in each disease categorization after running all tweets (1.2 million tweets) through the best classification model.

**Classification Performance (Addressing R2)**

In the first stage of the categorization process for the ground truth tweets, tweets were first classified as being relevant or not relevant to zika (Table 2).

Table 2. Sample tweets relevant and not relevant to zika.

| Classification | Sample Tweets |
| --- | --- |
| Relevant | I guess we can now add Zika to the list of STDs<br>Millions of GM mosquitoes to fight Zika virus in Caymans |
| Not Relevant | After violent crime, shoddy construction and police brutality in Brazil, Zika seems minor.<br>#MoreTrustedThanHillary going to Brazil during Zika virus season |

Tweets that were relevant were then categorized as being about symptoms, treatment, transmission, or prevention (Table 3).

Table 3. Sample tweets from the four categories of relevant tweets.

| Classification | Sample Tweets |
| --- | --- |
| Symptoms | WHO sees scientific consensus on Zika virus as cause for disorders<br>Puerto Rico Reports First Zika-Related Death Amid Outbreak |
| Treatment | Healthcare providers: See CDC guidelines for caring for pregnant women w/ possible #Zika exposure.<br>Zika drug breakthrough may be the first step towards treatment. |

| Transmission | Zika virus strain responsible for the outbreaks in Brazil has been detected in Africa<br>Zika threatens TWO BILLION people across the world: New maps reveal where virus is likely. |
|---|---|
| Prevention | #CDC officials say men who have the #Zika virus should wait 6 months before trying to conceive. #health<br>Senate Nears Deal for at Least $1.1 Billion to Fight Zika Virus |

One-thousand four hundred and sixty-seven tweets were manually labeled to train the classifiers and evaluate their performance. Figure 2 provides the distribution of the relevant tweets in the four categories. As seen from Figure 1, the distribution of the labeled gold standard dataset was similar to the distribution of the large data corpus.

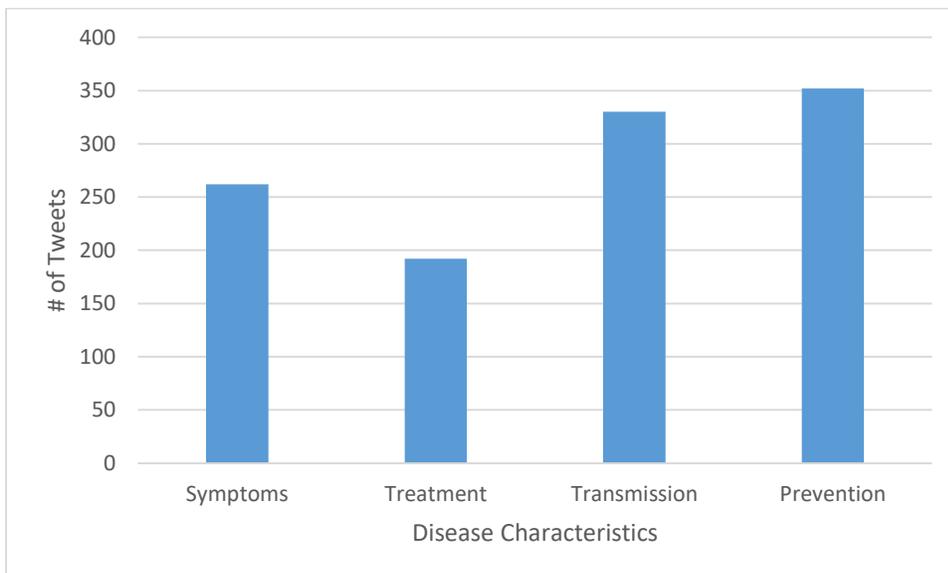

Figure 3. Number of tweets from the labeled dataset for each of the four categories of disease characteristics.

*Inter-rater Reliability*

Fleiss's Kappa was calculated for each category to check inter-rater reliability between three labelers with domain expertise in microbiology and immunology. Kappa values ranged from 0.62 to 0.93 for both the relevancy and the four disease characteristics. This indicates substantial to almost perfect agreement among the raters [20]. Once the inter-rater reliability was found to be substantial, a model needed to be built based on the gold standard dataset.

*Classification and Performance*

The table below gives the performance of different classifiers on 1,467 preprocessed Twitter data to find the relevancy of the tweet towards Zika (Table 4). Unigram features were extracted from the texts using the Weka toolbox [21]. For this dataset, the classifiers perform fairly well with AUC values ranging from 0.78 to 0.94. MNB outperforms other classifiers using both the metrics F Measure (0.86) and AUC (0.94) (Table 4). MNB classifiers perform better for data sets that have a large variance in document length (in this case, the length of the tweets) by incorporating the evidence of each appearing word into its model [22].

Table 4. Different classifier performances for detecting relevant tweets using J48, MNB, Bayesian networks (Bayes Net), SMO using SVM, as well as bagging or bootstrapping (Bagging) techniques.

| Classifier | TP | FP | Precision | Recall | F Measure | AUC |
| --- | --- | --- | --- | --- | --- | --- |
| J48 | 0.821 | 0.390 | 0.812 | 0.821 | 0.815 | 0.784 |
| MNB (bayes) | 0.880 | 0.368 | 0.881 | 0.880 | 0.868 | 0.943 |
| Bayes Net | 0.832 | 0.479 | 0.821 | 0.832 | 0.812 | 0.837 |
| SMO | 0.895 | 0.252 | 0.892 | 0.895 | 0.892 | 0.822 |
| Bagging | 0.857 | 0.411 | 0.852 | 0.857 | 0.843 | 0.877 |

The tables below detail the actual and predicted results of MNB using the confusion matrix (Table 5). The class imbalance was affecting the classifier performance. Although the AUC value was high (0.94), the classifier predicted a tweet was relevant more often than not relevant since ~80% of the tweets belong to the relevant category. This also affected the false positive rate which was much higher than the false negative rate as seen in Table 5.

Table 5. Confusion Matrix using the best classifier MNB.

| Predicted / Actual | 1 (relevant) | 0 (Not) |
| --- | --- | --- |
| 1 (relevant) | 1116 | 21 |
| 0 (Not) | 155 | 175 |

The table below gives the performance of different classifiers on 1,135 preprocessed twitter data to find the categorical classification (symptoms, treatment, transmission, and prevention) of the tweets (Table 6). Again, the classifiers performed fairly well with AUC values ranging from 0.83 to 0.94. With this dataset, MNB outperforms other classifiers again.

Table 6. Different classifier performances for detecting the four disease categories within the relevant tweets using J48, MNB, Bayesian networks (Bayes Net), SMO using SVM, as well as bagging or bootstrapping (Bagging) techniques.

| Classifier | TP | FP | Precision | Recall | F Measure | AUC |
| --- | --- | --- | --- | --- | --- | --- |
| J48 | 0.694 | 0.122 | 0.702 | 0.694 | 0.695 | 0.838 |
| MNB | 0.784 | 0.084 | 0.787 | 0.784 | 0.785 | 0.940 |
| Bayes Net | 0.697 | 0.121 | 0.729 | 0.697 | 0.702 | 0.885 |
| SMO (SVM) | 0.775 | 0.088 | 0.780 | 0.775 | 0.777 | 0.877 |
| Bagging | 0.727 | 0.112 | 0.741 | 0.727 | 0.730 | 0.901 |

In order to further understand the best classifier's performance a confusion matrix was created. The tables below detail the actual and predicted results using the MNB classifier (Table 7). Here, again the proportion of categories in the gold standard dataset affects the classifier predictions. The treatment was predicted the least number of times (169 out of 1135). However, the diagonal values (True positives) were higher than the misclassification values which accounts for the high AUC value. That said, there was also a noticeable overlap between Transmission and Prevention (42 tweets belonging to Prevention categorized as Transmission and 41 tweets belonging to Transmission categorized as Prevention). The reason these overlap may be due to the fact that the words mosquito and sex was used to describe how Zika is transmitted and how to prevent transmission. Also, they are closely linked in that prevention cannot occur unless the mode of transmission is known.

Table 7. Confusion Matrix using the best classifier MNB.

| Predicted ——————— Actual | Symptoms | Treatment | Transmission | Prevention |
|---|---|---|---|---|
| Symptoms | 205 | 4 | 45 | 8 |
| Treatment | 11 | 146 | 6 | 29 |
| Transmission | 20 | 4 | 264 | 41 |
| Prevention | 20 | 15 | 42 | 275 |

Based on the above results, the two-staged classifier system was able to have a high precision, as well as a high recall performance in categorizing the tweets into relevant and not relevant, and further classifying the tweets into the four disease categories. Once the performance of the model based on the gold standard data set was confirmed to have high precision and recall, the model needed to be tested on a new set of tweets.

*Confusion Matrix Error Analysis*

Similar to the study by Jiang [23], an error analysis was performed on the large test dataset by randomly selecting 530 tweets, manually labeling them, and comparing the results using the best performing MNB classifiers for identifying relevancy (Table 8), and then further categorizing the relevant tweets into the four subcategories (Table 9). Reevaluating the performance using the test set enables the unbiased assessment of the generalization error of the final chosen model. Here, 530 tweets were further annotated and tested the MNB classifiers from the earlier section. High precision and recall values were obtained for the relevance classifier with Precision =0.99 and Recall =0.99. Hence, the F measure was also 0.99 (harmonic mean of precision and recall). This high performance of the classifier indicates that the gold standard dataset was a good representation of the distribution of the tweets in the large data corpus. Moreover, the dataset was significantly less noisy when comparing the performance of these models with other domain areas such as gender based violence [5], drug-related tweets [24] and religion [25].

Table 8. Relevant tweet error analysis.

|  | Annotator |  |  |  |
|---|---|---|---|---|
|  |  | Not-relevant | Relevant |  |
| MNB | Not-relevant | 23 | 7 | 30 |
|  | Relevant | 7 | 493 | 500 |
|  |  | 30 | 500 | 530 |

Table 9. Disease characteristics error analysis results.

|  |  | Annotator |  |  |  |  |
|---|---|---|---|---|---|---|
|  |  | Symptoms | Treatment | Transmission | Prevention | Total |
| MNB | Symptoms | 88 | 0 | 5 | 1 | 94 |
|  | Treatment | 2 | 29 | 3 | 6 | 40 |
|  | Transmission | 13 | 1 | 129 | 25 | 168 |
|  | Prevention | 5 | 0 | 10 | 152 | 167 |
|  | Total | 108 | 30 | 147 | 184 | 469 |

Even though the classes were unbalanced, high precision and recall values were still obtained (Table 10). An overall high F measure of 0.9 was obtained. This further indicated the gold standard dataset was a good representation of the tweets, as well as the disease categories in the larger corpus and that the dataset was not very noisy.

Table 10. Precision recall and F measure for each of the four disease characteristics.

| Category | Symptoms | Treatment | Transmission | Prevention | Average |
|---|---|---|---|---|---|
| precision | 0.98 | 0.97 | 0.86 | 0.94 | 0.94 |
| recall | 0.81 | 0.97 | 0.88 | 0.83 | 0.87 |
| F | 0.89 | 0.97 | 0.87 | 0.88 | 0.90 |

The error analysis indicated that the classifiers performed well with the unseen test data and were generalized enough to work with the large dataset. The dataset was further examined with a focus on the insights provided in the disease categories. More specifically, the topics discussed on Twitter in the symptoms category was examined to discover the latent semantic topics discussed.

**Topical Analysis (Addressing R3):**

*Topic Modeling*

Below the results of LDA are discussed for each of the four disease characteristics (Tables 13-14). Based on heuristic results, the number of topics was restricted to 5 while discussing the topics for each category. Topic modeling results are shared here [26] for the research community to examine the outcome of using topic modeling, as well as the overlap among the topics generated. First the results for the three categories prevention, transmission and treatment will be discussed. Then a more detailed analysis of the topic modeling results for the symptoms category will be discussed along with the misinformation tweeted by users in that domain.

*Prevention, Transmission and Treatment*

Table 11 provides the topics for the three categories along with the sample tweets in each topic.

Prevention: Within the prevention topics, topic #1 was about the need to control/prevent spread, topic #2 was about the need for money to combat mosquitoes and research treatments, topic #3 was about ways to actually prevent spread, topic #4 was about the bill to get funds, and topic #5 was research (Table 11). These topics were not surprising considering all the discussion about how to prevent Zika, the need for funding to prevent Zika, and the research required to find a cure for Zika since it is an emerging disease. There is also a need to better understand Zika virus, the disease it causes, and ways to combat it [27].

Transmission: In transmission, there was a strong overlap in topics #1 (vector i.e. mosquitoes for Zika) and #4 (disease spread) that highlight the overlap between the mosquito spread, and the disease in general. Looking at the tweets, both highlight the concerns and risks associated with zika spread. Another topic (#2) was sexual which is another mode of transmission besides through mosquitoes. The next topic (#3) was infants, who are most affected by this epidemic due to the risk of microcephaly. The final topic (#5) was sports since the tweets were collected during baseball season just before the olympics and many athletes were concerned about getting affected with Zika while competing in Rio, 2016.

Treatment: There was slight overlap between topics #1 (lack of treatment) and #3 (vaccine development) for treatment primarily due to the large co-occurrence of the word vaccine in both these topics. Blood testing (#4) was another major topic since some people got infected with Zika after receiving a blood transfusion. Since no treatment exists, a lot of research is focused on developing a drug for Zika, which is why test development (#5) was the final topic.

Table 11. Prevention, transmission, and treatment topic modelling results.

| Disease Characteristic | Topic | Sample Tweets for each Topic |
|---|---|---|
| Prevention | (#1) Control | RT @DrFriedenCDC: A2. The best way to prevent #Zika & other diseases spread by mosquitoes is to protect yourself from mosquito bites. #Reut |
| | (#2) Money Need | #healthy Congress has not yet acted on Obama's $2 billion in emergency funding for Zika, submitted in February |
| | (#3) Prevention | RT @bmj_latest: Couples at risk from exposure to Zika virus should consider delaying pregnancy, says @CDCgov |
| | (#4) Bill | https://t.co/Ke12LOdypf Senate Approves $1.1 Billion In Funding To Fight The Zika Virus #NYCnowApp |
| | (#5) Research | Florida is among those at greatest risk for Zika. @FLGovScott's sweeping abortion bill blocks scientists' access to conduct research |
| Transmission | (#1) Vectors (mosquitoes) | This map shows the Northeast is at risk for Zika mosquitos this summer |
| | (#2) Sexual | @user1 First Sexually Transmitted Case Of Zika Virus In U.S. Confirmed |
| | (#3) Infants | CDC reports 157 cases of U.S. pregnant women infected with Zika virus. |
| | (#4) Spread | Zika strain from Americas outbreak spreads in Africa for first time: WHO (Update) |
| | (#5) Sports | MLB moves games from Puerto Rico due to Zika concerns....uh..what about the Olympics?? Can't be good. |
| Treatment | (#1) Lack of Treatment | RT @DrFriedenCDC: Much is still unknown about #Zika and there is no current medicine for treatment or vaccine to prevent the virus. |
| | (#2) Zika Test | Rapid Zika Test Is Introduced by Researchers The test, done with a piece of paper that changes color if the virus … |
| | (#3) Vaccine Development | Researchers discover structure of Zika virus, a key discovery in development of antiviral treatments and vaccines |
| | (#4) Blood Test | Experimental blood test for Zika screening approved |

| | (#5) Test Development | New mouse model leads way for #Zika drug, vaccine tests |

Symptoms: In the topic model results for symptoms, topics #1 (zika effects), #2 (brain defects) and #4 (zika scarier than thought) were well separated while topics #3 (confirmation of defects) and #5 (initial reports) overlap significantly (Figure 4 & Table 12). The topics are described in Table 16. Topics #3 and #5 overlap for symptoms because a lot of the initial reports were about health official confirming defects. Topic #3 was more about the defects that were confirmed while topic #5 focused on where reports came from.

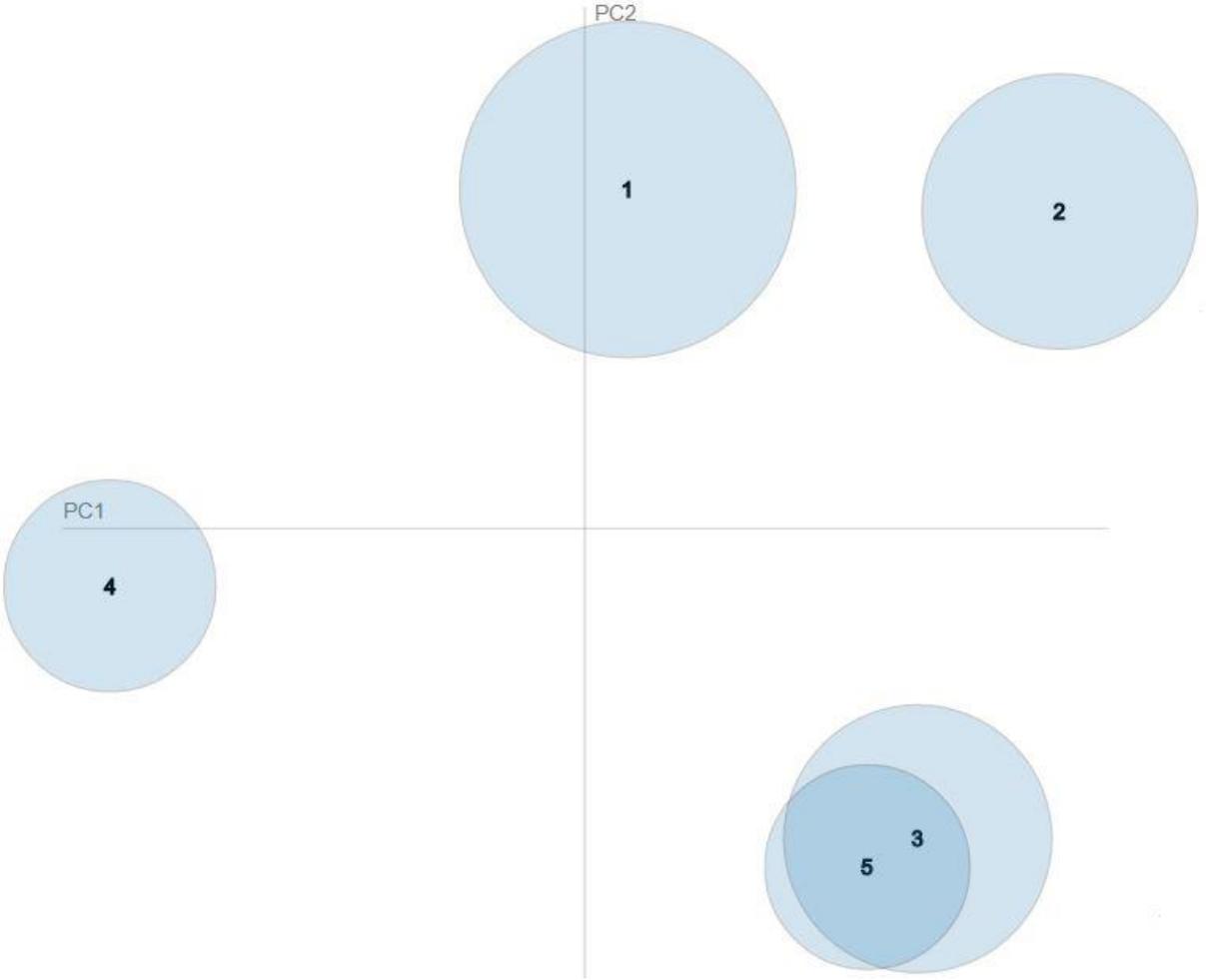

Figure 4. LDA of symptoms.

All three topics (#1, #3, #5) were closely related in that they are about the defects caused by Zika, but point to slightly different concepts (Table 12). For example, microcephaly is not the only defect, there is also Guillain-Barre which would be topic #1. Topic #2 focuses on microcephaly because that was a major topic of discussion as seen in these tweets: *"RT*

*@USATODAYhealth: Zika affects babies even in later stages of pregnancy. Microcephaly seen in babies from moms infected in 6th month"* and *"Zika Virus May Cause Microcephaly by Hijacking Human Immune MoleculeFetal brain model provides first clues on how Z…"* Topic #3 contains tweets when the defects were confirmed to be caused by Zika and not something else: *"Zika linked to fetal brain damage: Finnish study: infectious Zika virus from fetal tissue in cell culture.The virus"*, and *"Enough conspiracy theories; nature is nasty enough: U.S. health officials confirm Zika cause of severe birth defects"*.

Topic #4 for symptoms was primarily just people discussing a British Broadcasting Corporation article [28] on how more birth defects have been linked to Zika and that the virus was expected to travel further than initially thought, leading to experts saying Zika is scarier than they first thought. The statement of Zika being "scarier than we first thought" by the CDC was a big topic on twitter *"CDC says zika virus scarier than thought as US prepares for outbreak: On Monday, the U.S. Centers for Disease…"*, and this also affected the U.S political environment: *"#2016elections U.S. Officials Warn Zika Scarier Than Initially Thought: By Timothy Gardner and Jeff Mason WA…"*, , This discussion lead to some more tweets about the danger of Zika virus: *"The Edge: Zika Is Now Even More Terrifying"*, *"Zika virus 'shrinks brains' in tests"*, and *"#Zika Survivor Says 'I Could Feel My Skin Shrinking' CBS Boston"*. These tweets demonstrate how a statement by the CDC can be spread and how users can tweak the wording of these CDC statements to generate more concern than the actual impact of the disease. Finally, topic #5 includes tweets about initial reports of Zika outbreaks and deaths.

Misinformation: Within symptoms, several tweets in topic #1 were calling Zika a hoax, *"Zika HOAX exposed by South American doctors: Brain deformations caused by larvicide chemical"*, *"The Zika Virus is a hoax! It is like calling the common cold an epidemic. It's what they put in the drinking water."*, and *"CDC likely fabricating link between Zika virus and microcephaly cases"*. However the CDC has stated multiple times that Zika and microcephaly are definitely linked *"CDC: Zika definitely causes severe birth defects"*, and *"Here's a #Zika basic: Zika infection during pregnancy can cause some severe birth defects"*. Some of the people saying Zika is a hoax are misunderstanding this quote from the CDC *"People usually don't get sick enough to go to the hospital, and they very rarely die of Zika. For this reason, many people might not realize they have been infected. Once a person has been infected, he or she is likely protected from future infections"*. This statement is true for the majority of healthy adults. However, for infants it can cause microcephaly and in some cases Guillain-Barre syndrome in healthy adults *"Symptoms of Guillain-Barre syndrome include weaknesses in arms & legs. GBS is linked w/ #Zika"*. There also has already been a couple of deaths due to Zika as was detailed in topic #5 in the LDA. The CDC has also been directly answering people's questions about Zika on twitter. One user tweeted at the CDC *"Why is #Zika of particular concern to women who are pregnant or considering becoming pregnant? #WellnessWed #ZikaVirus"* to which the CDC responded *"Zika infection in pregnancy can cause microcephaly and other severe brain defects. http://1.usa.gov/1Pf79sK #WellnessWed"*. This shows that while some misinformation is still getting tweeted, the CDC is working to get the correct information out there. This is useful because it shows the CDC could potentially target specific user groups directly through the topic modelling approach, and respond to users within a topic group with a similar response that can allow the information to spread to a larger population with relatively lesser effort.

Table 12. Symptoms topic modelling results.

| Topic | Words | Tweets |
| --- | --- | --- |

| (#1) Zika Effects | infect, babies, mosquito, cause, microcephaly, symptom, pregnancy | RT @USATODAYhealth: Zika affects babies even in later stages of pregnancy. Microcephaly seen in babies from moms infected in 6th month |
|---|---|---|
| (#2) Brain Defects | brain, link, studies, microcephaly, baby, disorder, cause, damage, infect, fetal | 'Zika Virus May Cause Microcephaly by Hijacking Human Immune MoleculeFetal brain model provides first clues on how Z… |
| (#3) Confirmed Defects | defect, cause, birth, confirm, health, severe, link, official | Enough conspiracy theories; nature is nasty enough: U.S. health officials confirm Zika cause of severe birth defects |
| (#4) Scarier Than Thought | scarier, than, thought, us, official, health, CDC, warn, learn, first | #breakingnews Zika Virus 'Scarier Than We First Thought,' Warn US Health Officials |
| (#5) Initial Reports | first, report, death, case, puerto, confirm, rico, cause, colombia, defect | Colombia Reports First Cases of Microcephaly Linked to Zika Virus - Sun Jan 09 15:13:20 EST |

In this section, the topic modeling results generate insightful results that allow researchers to understand the citizens' concerns, as well as the misinformation spread. According to the theory of LDA, each topic represents certain common properties, which reflects the pattern in the tweets. Finding out the exact meanings of the topics requires additional information and domain knowledge. We see that for each of the disease characteristics, the discovered topics can be easily interpreted by using some domain specific knowledge.

# Discussion
**Classification Analysis**

One of the interesting findings of our analysis was the fact that the Multivariate Naive Bayes classifier or the MNB outperformed the other more popular classifiers in text analytics: random forest (J48) and SVM. According to one study [29], this has to do with the class imbalance issue in our dataset: both for the relevancy case, as well as for the disease categories case. This also highlights the possible orthogonality of the features used in our study: the unigrams. Specifically in this dataset, measuring the likelihood of the features in a given class independently outperforms other complex models such as J48 and SVM. This possibly also relates to the fact that the data are less noisy since they have been evaluated by expert annotators. Naive bayes is one of the simplest classification models available to us, but it is nonetheless among the most effective for this dataset. This result is non-intuitive, but is not surprising when we consider that using text for classification is relatively imprecise compared to other types of data. In datasets with large amounts of error, simpler models are less likely to overfit the data. Hence we recommend that future research on text analytics begin with Naive Bayes and then proceed to using more complex models to see if these actually improve classification accuracy.

**Annotation Observations**

One major issue annotating tweets was what to do about news tweets like this one: "*Your Wednesday Briefing: Bernie Sanders, Hillary Clinton, Zika Virus: Here's what you need to know to start ...*". The issue was that this does give information about Zika in that it tells what news stations were talking about it and what else was going on at the same time as the Zika outbreak. However, the tweet itself does not give any information about Zika symptoms, treatment, transmission, and prevention. The researchers decided to code these tweets as relevant because they were about Zika, but not include them in the disease characteristics annotations since they do not have any information about Zikas disease characteristics. Sporting events in general were also included because they could be sources of transmission from athletes and fans not taking proper precautions.

**Topic Modeling**

All of the topics under the different disease characteristics fit the characteristic. For example, control, money need, prevention, bill, and research were all major topics of prevention discussions. This indicates the classification model accurately labelled tweets. It also indicates tweets about major topics were collected and accurately reflected in our topic model. Also, while all four disease characteristics are important, symptoms was discussed in detail because the researchers felt it included the most important information for public health officials to know especially once the misconceptions/ misinformation about Zika being a hoax was found. Categorizing the symptoms into the different topics using topic modeling also allowed us to get deeper into the themes within the symptoms category that can allow a more targeted interaction with agencies like CDC and specific users to provide interventions for misinformation spread.

# Limitations

While we have clearly provided an exploratory framework in mining the different characteristics of Zika, we point out the limitations we face in our dataset, and the use of social media.

*Language constraint:* We have restricted our study to English language tweets that certainly limit the strength of our study. This is more critical to address given that South American countries were initially affected by Zika. This also restricts our analysis of measuring disease outbreak which is why we refrained from doing so in our study.

*Keyword constraint:* As described in the Data collection section, we used the keywords *Zika, Zika virus, Zika treatment,* and *Zika virus treatment* in our study. One interesting observation here is that although the keyword "treatment" was part of the crawling process, the treatment subcategory was still the smallest class in the distribution of the dataset (see Figures 2 and 3). Clearly, this certainly affected our dataset. This also would overlook tweets that referred to the disease in a different name or talked about the disease without using the word Zika.

*Gender & Polarity constraint:* Only 49% were labeled by the gender API using the profile name (Table 1). Similarly, around 10% of the tweets were not labeled on their polarity. This needs to be addressed moving forward with this study by creating a customized gender recognition tool using machine learning specifically for twitter data.

# Conclusion and Future Work

This is one of the first studies to report successful creation of an automated content classification tool to analyze Zika–related tweets; specifically in the area of epidemiology. Such a system will help advance the field's technological and methodological capabilities to harness social media sources for disease surveillance research. Future studies will include creating an automated technique to detect misinformation using tweets to allow for well-targeted, timely interventions. Such a platform will generate data on emerging temporal trends for more timely interventions and policy responses to misinformation on Zika studies leveraging information sources including blogs, news articles, as well as social media data.

**Acknowledgements**
We would like to thank Dr. Rua, Scott Holdgreve, Ryan, Becker, Amber Todd, and Sampath Gogineni for their help.

**Abbreviations**
LDA: Latent dirichlet allocation
CDC: Center for Disease Control
WHO: World Health Organization
GBS: Guillain-Barre syndrome
DC: Disease classifier
RC: Relevancy classifier
API: Application program interface
SVM: Support vector machine
J48: Decision tree
MNB: Multinomial naive bayes
SMO: Sequential minimal optimization
AUC: Area under the curve